# Competing interaction in magnets: the root of ordered disorder or only frustration?

Per Nordblad
Uppsala University, Department of Engineering Sciences, Box 534, SE-751 21 Uppsala

Abstract
What does the equilibrium atomic, molecular or spin configuration of a glass phase look like? Is there only one unique equilibrium configuration or are there infinitely many configurations of equal energy? The processes and mechanisms governing the path towards equilibrium, i.e. the dynamics of glassy systems, provide insights to these questions. Here we discuss the intrinsic dynamics of different glassy magnets: of spin-glasses, frustrated ferromagnets, superspin-glasses and other nanostructured systems with competing ferro- and antiferromagnetic interactions and randomness in their spatial distribution.
This paper is intended as a brief update on some unsolved problems and the current empirical status in the field of disordered and frustrated magnetism.

## 1. Introduction

Temperature (thermal energy) favours disordered states of matter. Interatomic forces strive to condense matter to states of perfect order. Randomness and defects – quenched disorder necessarily accompany the atomic structural order of crystalline matter. Magnetic materials mirror and illustrate consequences of these counteracting influences. Strong direct exchange interaction between the atoms in iron causes ferromagnetic order well above room temperature. The degree of atomic disorder in the crystal structure governs the magnetic functionality of iron-based alloys – allowing soft (transformer sheet) or hard (permanent magnet) magnetic behaviour.

Certain alloys and compounds possess more complex interaction patterns between the magnetic atoms than iron; in some materials competing ferro- and antiferromagnetic interactions occur which give rise to frustration conveying conflicting information to the atomic magnetic moments on the direction to point in and low temperature states with disordered order: Spin-glasses. Different natural and artificial nanostructured materials possess magnetic properties that cover magnetic phenomena from pure ferromagnetic to subtle glassy behaviour; making the analyses of empirical observations challenging.

## 2. Magnetic hysteresis

Exchange interaction is the source of magnetic order. Disorder and anisotropy make the low temperature ordered state hysteretic in temperature (T) and magnetic field (H). Much insight, but also confusion, about intrinsic and extrinsic magnetism can be obtained from standard measurements of hysteresis loops (magnetization (M) vs. magnetic field loops) and M vs. T sweeps in zero-field cooled (ZFC) and field cooled (FC) protocols [1]. The existence of hysteresis implies that the system occupies metastable spin (microscopic magnetic moment) configurations that, when kept at constant T and H, have to reorganize their spin or domain structure to attain thermodynamic equilibrium. This reorganization may occur on any time scale ranging from the microscopic relaxation time of the intrinsic magnetic entities through the observation time of an actual experiment to geological timescales (infinitely long in theoretical physics terms).

## 3. Glassy dynamics – aging – magnetic model systems

Ideal crystalline materials are in thermodynamic equilibrium states. In contrast, glassy structures are non-equilibrium states that spontaneously and eternally re-organize their atomic configuration. As a consequence, in addition to response functions on extended dynamic ranges, glasses have intrinsic age dependent physical properties. In magnetism, ferro- and antiferromagnetically ordered phases represent equilibrium states, whereas spin-glasses are non-equilibrium systems. Equal strength and equal amount of ferro- and antiferromagnetic interaction and randomness in the interaction pattern cause **disorder** and **frustration**; the two physical ingredients required to create spin-glasses.

The dilute magnetic alloys Cu(Mn) and Au(Fe), and the mixed magnetic ion compound $Fe_{0.5}Mn_{0.5}TiO_3$ are extensively studied archetypal spin-glasses [2]. The origin of disorder and frustration are different in these systems. In a dilute alloy, the magnetic ions are randomly distributed in a crystalline non-magnetic metallic matrix and the distance dependent oscillatory electron mediated RKKY-interaction [3] gives rise to random distribution of ferro- and antiferromagnetic interaction providing disorder and frustration in the system. Iron and manganese ions in $Fe_{0.5}Mn_{0.5}TiO_3$ form hexagonal planes where the intra-plane interaction is ferromagnetic between Fe ions and antiferromagnetic between Mn ions, and the inter-plane interaction is antiferromagnetic for both ions [4]. The random distributions of Fe and Mn atoms on the metal sites in the crystal structure cause equal amounts and strengths of ferro- and antiferromagnetic interaction and give rise to disorder and frustration and thus spin-glass behaviour at low temperatures (see Figs. 1 and 2).

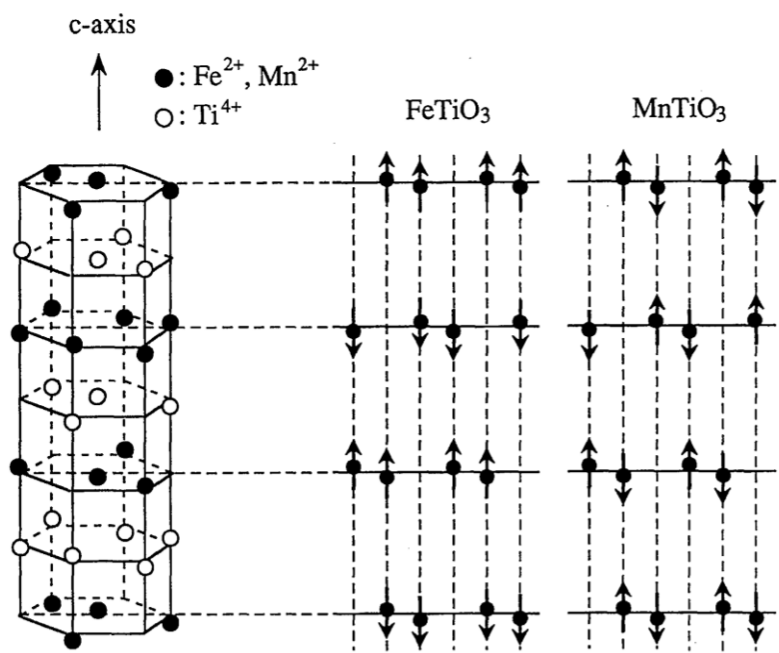

Fig. 1 Structure of the parent compounds $FeTiO_3$ and $MnTiO_3$ of the mixed system $Fe_xMn_{1-x}TiO_3$; from ref. [4].

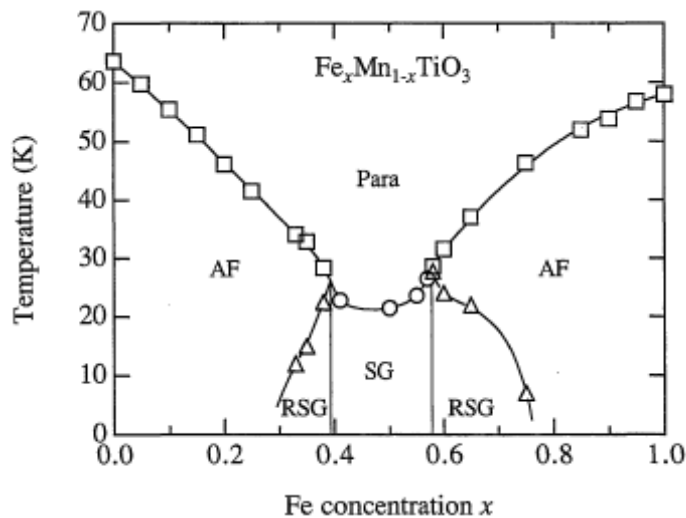

*Fig. 2 T-x phase diagram of $Fe_xMn_{1-x}TiO_3$. AF antiferromagnetic, SG spin-glass and RSG re-entrant spin-glass phases; from ref. [4].*

Dynamics on extended time scales characterise the mechanical and dielectric properties of glasses. The magnetic response of the low temperature spin-glass extends over time scales ranging from the atomic spin flip time ($\tau_0 \sim 10^{-13}$ s) to geological times, limited only by the geometric extension of the material. In addition, the response function in spin-glasses depends on the age of the system owing to spontaneous re-organisation of the frustrated spin structure.

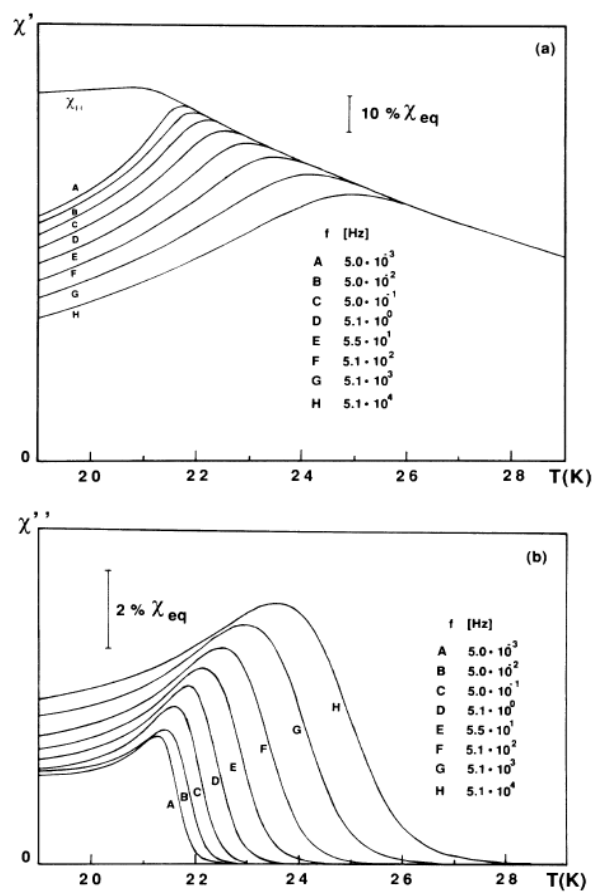

*Fig. 3 Low field ac-susceptibility measured at frequencies $5\ 10^{-3} - 5.1\ 10^4$ Hz of $Fe_{0.5}Mn_{0.5}TiO_3$; from ref. [5].*

Critical fluctuations on approaching a second order phase transition occur on a length scale that diverges at the transition temperature. Associated with the diverging correlation length is a diverging relaxation time – critical slowing down – according to:

$$\xi \sim |t|^{-\nu} \text{ and } \tau \sim \xi^z \sim |t|^{-z\nu}, \tag{1}$$

where $\xi$ is the correlation length, t (in this equation) is the reduced temperature: $t=(T-T_c)/T_c$, $\nu$ and z are critical exponents, and $\tau$ is the relaxation time. This also implies that close to a second order phase transition, an experimental probe with a given observation time $t_{obs} < \tau$, observes fluctuations up to a certain length scale $L < \xi$. Observation of critical slowing requires that the magnitude of the applied field is weak enough to yield a linear response of the magnetisation i.e. M/H has to be independent of the field magnitude (implying that no field-driven processes are created). Fig. 3 shows the in and out-of-phase component of the low field ac-susceptibility of $Fe_{0.5}Mn_{0.5}TiO_3$ over a wide frequency range near the spin-glass transition. Analyses of these results in terms of Eq. 1 yield $z\nu = 10\pm0.5$ [5] which can be used as a reference value for the dynamic exponent of three-dimensional Ising spin-glasses [6].

The spontaneous re-organisation processes of the spin structure in a spin-glass (aging) occur on longer and longer length scales including increasingly large volumes with time at constant temperature [7-9]. These processes are mirrored in low field magnetic relaxation experiments. Fig. 4 shows zero field cooled magnetic relaxation experiments on a Cu(Mn) spin-glass at a temperature below $T_g$ [8]. The relaxation of the magnetization develops with observation time (t) in a logarithmic fashion and is strongly dependent on the time spent at constant temperature ($t_w$) before the probing magnetic field is applied. All M vs. log t curves show marked inflection points at timescales of order $t_w$. The curves measured at the longest relaxation times assume age independent levels at short observation times, indicating that the system first reaches equilibrium configurations on short length scales. Application of excessive magnetic fields causes non-linear response and introduces field-driven processes [9-10].

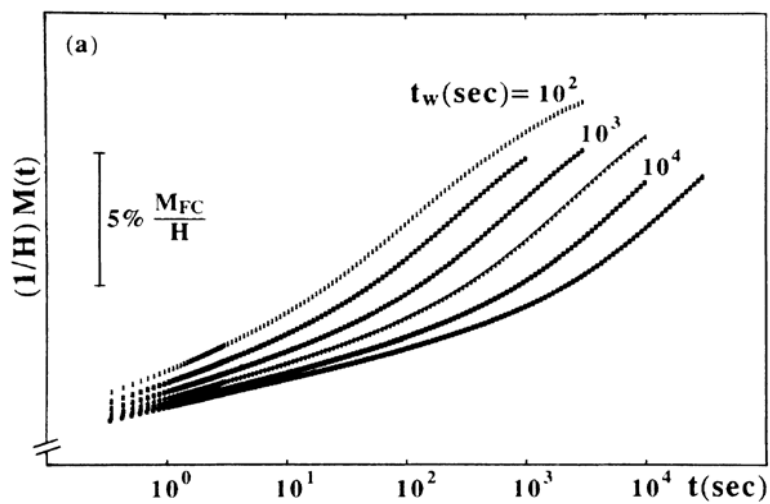

Fig. 4 *M vs. log(t) curves measured after different waiting times at constant temperature at $T/T_g=0.8$ in a Cu(Mn) spin-glass illustrating the aging phenomenon; from ref. [8].*

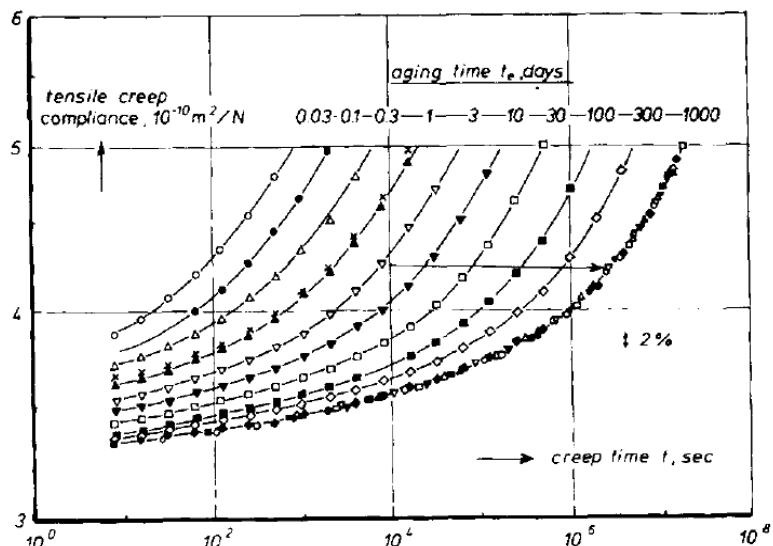

*Fig. 5 Tensile creep vs. log(t) measured after different waiting times at constant temperature below the glass temperature of PVC; from ref. [11].*

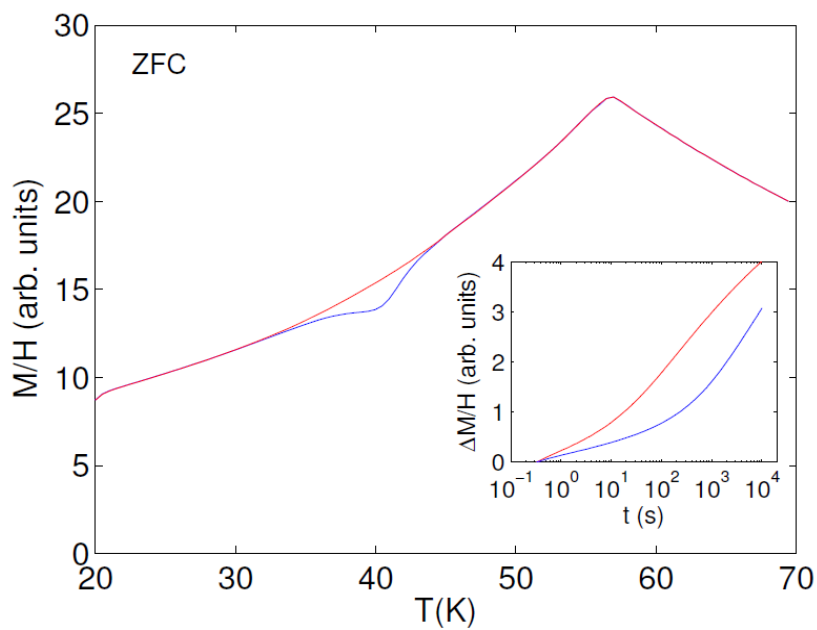

*Fig. 6  M/H vs. temperature for a Cu(Mn) spin-glass. The cusp at about 60 K marks the spin-glass character. The sample has been cooled in zero magnetic field to 20 K where a weak magnetic field is applied and the susceptibility is recorded on heating at the same rate for the two recorded curves. The curve without a dip is recorded after continuous cooling in zero field to 20 K, whereas the curve with dip is recorded after continuous cooling in zero field including a halt at 40 K for 3000 s. The inset shows ZFC-magnetic relaxation curves measured at 40 K after two different waiting times ($t_w$ = 0 s (red) and $t_w$ = 3000 s (blue)) before applying the magnetic field. The difference between these two curves at a time of 30 s is about the same as the difference between the two M vs T curves in the main frame at 40 K.*

The mechanical property - tensile creep - of PVC, illustrated in Fig. 5, shows a similar behaviour to spin-glasses. The aging behaviour in various structural glasses had been amply investigated long before the characteristics of aging in spin-glasses were discovered [11].

A counterintuitive property of spin-glasses is a marked memory phenomenon [9, 12-14]. If the spin-glass is kept at a constant temperature below the spin-glass temperature ($T_g$), spin correlations develop and slow down the response function as illustrated in Fig. 4. If the sample is cooled from this temperature to a lower one, the response function at this lower temperature appears unaffected by the ageing at the higher temperature, i.e. the response is the same as if the sample **had been cooled from above $T_g$ without a pause**. The spin-glass is rejuvenated - the spin structure that has developed at the first ageing temperature is irrelevant at a lower one. This indicates that the ultimate equilibrium spin-glass structure is chaotic in the sense that the equilibrium spin configuration at one temperature is different from the equilibrium configuration at any other temperature in the spin-glass phase. However, the spin structure that develops over time at constant temperature survives successive temperature cyclings to lower temperatures; i.e. re-heating the spin-glass to the temperature where the spin-glass was kept a constant temperature results in a spin-glass with a memory of the aging at this temperature during the initial cooling down. Fig. 6 illustrates the memory phenomenon using low field ZFC magnetization vs. temperature measurements. The inset shows ZFC magnetic relaxation measurements (M vs. log t experiments (similar to those in Fig. 4)) at the same temperature as the memory was imprinted, applying the same field as in the memory experiment; one relaxation curve is measured immediately after reaching the temperature and the other curve measured after waiting 3000 s at this temperature. Looking at the two curves at an observation time of 30 s, one observes that the difference between the two curves is about as large as the difference between the reference and memory curves in the main frame.

The results in Figs. 3, 4 and 6 summarize the key characteristics of the zero-field (non-field-driven) spin-glass phase: critical slowing down, relaxation over extended time scales below $T_g$, aging, rejuvenation (i.e. the chaotic nature of the spin structure on changing the temperature) and a memory of an imprinted spin structure on re-heating [9, 13, 15].

**4. Modelling and predicted phase diagrams**
Spin-glass modelling starts from the Hamiltonian: $H = -\Sigma J_{ij}S_iS_j$; where the geometry of the system and the distribution of the exchange interaction, $J_{ij}$, determine the level of realism. Theoretical breakthroughs are attributable to: Edwards and Anderson [16] (who invented the EA spin-glass order parameter), Sherrington and Kirkpatrick [17] (for the SK spin-glass model), Giorgio Parisi [18] (who determined an analytic solution to a mean field spin-glass model, still the only analytic solution to the spin-glass problem), Almeida-Thouless [19] (for the prediction within a mean field model of a line in the H-T plane that separates the paramagnetic from the spin-glass phase: the AT-line) and Fisher-Huse [7] (for the development of a droplet scaling model of equilibrium and non-equilibrium dynamics in spin-glasses). However, there remain gaps between experimental phenomenology and predictions from analytic modelling. On the other hand, recent massive Monte Carlo simulations ($10^{12}$ MC-steps) of 3D Ising spin-glass models [20] make direct comparisons between relaxation functions and phenomena observed on ideal model systems and real spin-glass materials possible, since these provide a region of overlapping time scales between simulations and experiments corresponding to $10^8 – 10^{12}$ MC-steps or $\tau_0$. Also, recent MC-simulations by Larson et al. [21] indicate that there is no AT-line in three dimensional spin glasses.

The predicted phase diagram of a frustrated magnet depends on the dimensionalities of the system and the distribution of the exchange interaction $J_{ij}$ – a spin-glass has zero mean. Skewed distributions of $J_{ij}$ give rise to re-entrant systems with transition sequences on lowering the temperature: paramagnetic - ferromagnetic (antiferromagnetic) - re-entrant states. The sequence: disordered (paramagnetic) - ordered (ferromagnetic) – disordered (spin-glass) at least word-wise describes a thermodynamic inconsistency. However, the ferromagnetic (antiferromagnetic) phase is a frustrated ordered state with complex dynamic properties; and represents a less ordered state than the low temperature spin-glass. A significant and related prediction is attributable to Griffiths [22] and concerns the behaviour of magnetically diluted Ising ferromagnets at temperatures below the Curie temperature of the non-diluted system. This concept, the Griffiths phase, is often employed to interpret results on complex magnetic oxides that, in addition, exhibit spin-glass or re-entrant spin-glass-like features.

**5. Re-entrant spin-glass**
Re-entrant ferromagnets have striking properties and their magnetic response is exceedingly sensitive to the magnitude of the applied magnetic field [23]. Thus, it is a demanding task to reach a regime of linear response to be able to investigate intrinsic (not field-driven) response functions. The low temperature spin-glass state of these systems exhibits a substantially enhanced field regime of linear response compared to the high temperature ferromagnetic phase and exhibits the dynamic characteristics of a pure spin-glass state. In contrast, the frustrated ferromagnetic phase at higher temperature possesses an enhanced ageing behaviour; the response function changes dramatically with age ($t_w$). The dynamics of the ferromagnetic phase show rejuvenation on decreasing temperature, but no memory phenomenon [24]. Thus, there are manifestations of disorder and frustration in the dynamics of the ferromagnetic phase of re-entrant systems; however, the equilibrium spin structures developed are more fragile and susceptible to temperature and field disturbances than the corresponding features in standard spin-glasses.

**6. Magnetic nanoparticle systems**

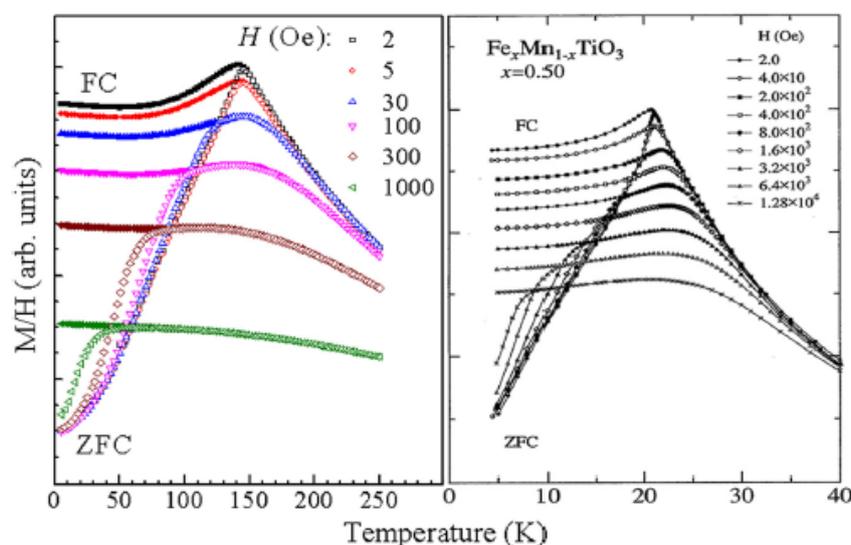

*Figure 7. ZFC- and FC-Magnetisation vs. temperature at different applied magnetic fields (as indicated in the figures). Left panel a dense nanoparticle system, from ref. [25] and right panel the Ising spin-glass $Fe_{0.5}Mn_{0.5}TiO_3$; from ref. [4].*

Superspins ( > 100 $\mu_B$) are formed in magnetically ordered nanoparticles – whether ferromagnetic, ferrimagnetic or antiferromagnetic (excess moments at the surface). The magnitude of the magnetic moment depends on the size of particles. The interaction between particles in an assembly can be purely dipolar if the particles are suspended in an insulating matrix that prohibits direct coupling by maintaining a distance between the surfaces of adjacent particles. The distance between particles determines the coupling strength. The individual relaxation time, $\tau$, of the particles is governed by its magnetic anisotropy energy, $E_a$, and follows the Arrhenius law: $\tau = \tau_0 e^{E_a/k_B T}$, where $E_a = KV$; K is the anisotropy constant,

V the volume of the particle, T is the temperature and $\tau_0$ the microscopic relaxation time of the particle at high temperature. In negligibly interacting (very dilute) particle systems – superparamagnets - the temperature dependence of the low field equilibrium susceptibility is governed by the Curie law. However at low enough temperatures, when $\tau$, becomes of the order of the observation time of the experimental probe, the particles appear blocked - the ac-susceptibility shows a frequency dependent maximum and the ZFC magnetisation vs. T exhibits a broad maximum and bifurcates from the FC magnetisation, which at lower temperatures deviates from the 1/T behaviour. The size distribution of the particles broadens the maximum of the ZFC susceptibility since the relaxation time depends exponentially on the particle volume.

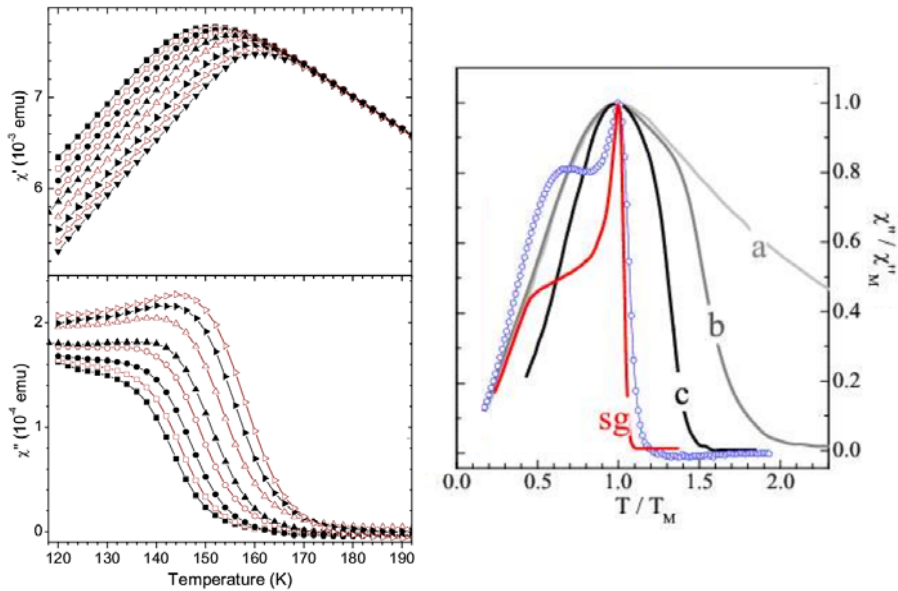

*Fig. 8 (Left panel) In- and out-of-phase components of the low field ac-susceptibility at 0.1 – 600 Hz of a monodispersed randomly closed packed (RCP) nanoparticle system [25]. Critical slowing down analysis yields zv=9.5. (Right panel) Comparisons of the out-of-phase component on approaching $T_g$. for the RCP sample, a model spin-glass [26] and results earlier studies of interacting nanoparticle systems (a) [27], (b) [28] and ( c) [29]; adapted from refs. [25].*

Figs. 7 and 8 show results from ZFC- and FC- magnetisation measurements at different fields [25, 4] and low field ac-susceptibility measurements at different frequencies on a closely monodispersed densely packed dipolarly interacting superspin-glass [25] together with corresponding data on an atomic Ising spin glass [26] and some strongly interacting nanoparticle systems [27-29]. Dense dipolarly interacting nanoparticle systems with narrow size distributions exhibit all the features of model spin-glasses, these being critical slowing down, aging, rejuvenation and memory, thus earning them the label super spin-glasses [30,31]. Systems with larger size distribution (implying vastly different relaxation time in-between smaller and larger particles) and/or more diluted systems show less stringent behaviour and a smooth transition from high temperature superparamagnetic behaviour to a frustrated low temperature state. Such systems form complex phase diagrams including closely superparamagnetic states with low particle concentrations and superferromagnetic states with high concentrations [32].

**7. Nanoscale phase separated magnets and multi-glasses**
Physically nanostructured materials, including magnetic clusters or grains and materials with intrinsic phase separation between ferro- and paramagnetic nano/micro-volumes exhibit magnetic properties governed by competing interaction. The magnetic properties of some of these systems resemble those of atomic spin-glasses or re-entrant spin-glasses. The number of governing parameters is, however, enhanced in comparison with atomic systems. The intergrain interaction patterns are often complex, ranging from only dipolar interaction in insulating matrices via direct exchange and RKKY-like mediated interaction in metallic matrices to superexchange and double exchange mechanisms in strongly correlated oxides.

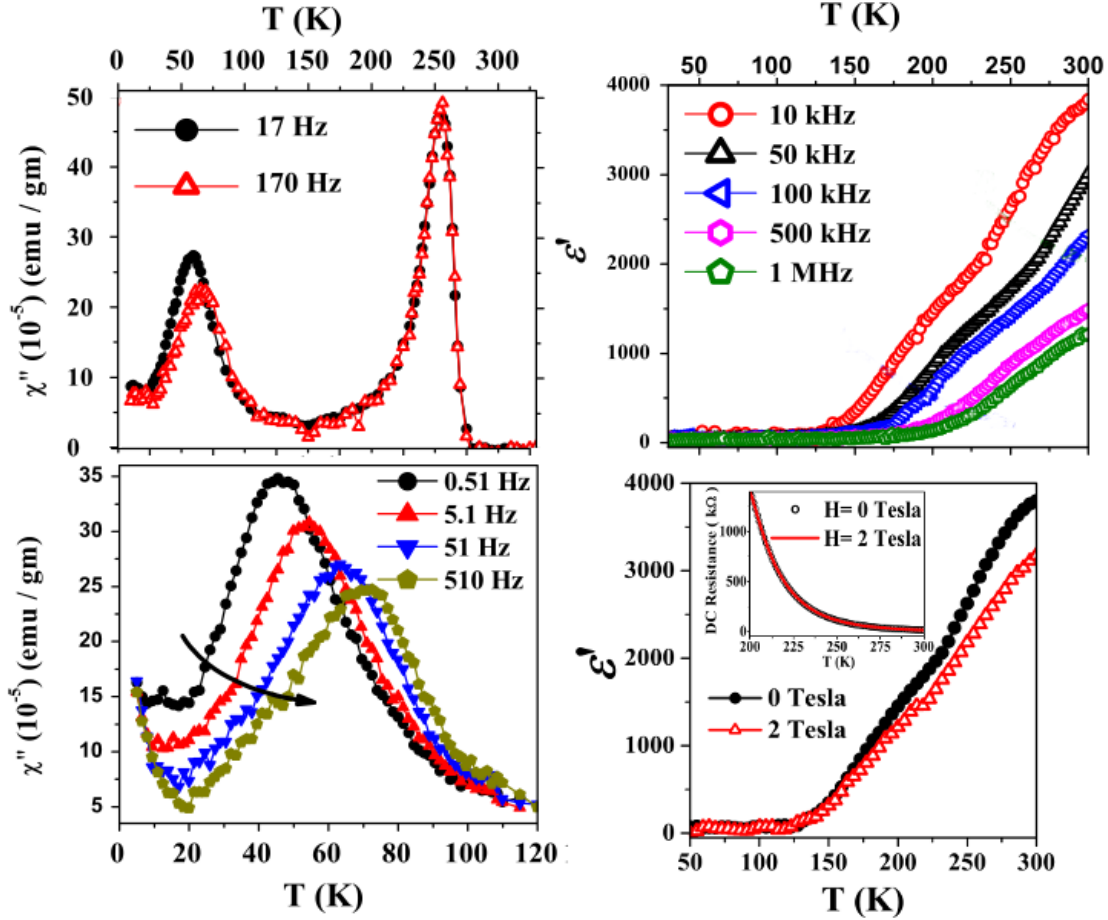

*Fig. 9 Ac-susceptibility ($\chi''$) (left panels) vs. temperature at different frequencies and dielectric constant $\varepsilon'$ (right panels) vs. temperature at different frequencies of $La_2NiMnO_6$. In the lower right panel, the main frame shows $\varepsilon'$ (at 10 kHz) measured in zero and 2 T fields, and the inset shows the resistivity of the sample at the same applied fields; adapted from Ref. [33].*

Electric manipulation of magnetic functionality is a long desired possibility that boosts the interest in multiferroics – materials that combine ferromagnetic with ferroelectric order. However, there are few examples of systems with such concomitant spin and dipole ordering, as well as strong intrinsic magneto-electric coupling. Recently, re-entrant spin-glass behaviour combined with dielectric glassiness including strong magneto-dielectric coupling was reported by D. Choudhury et al. [33]. These authors investigated a partly disordered double perovskite $La_2NiMnO_6$ that enters a frustrated ferromagnetic phase ($T_c$=270 K) and undergoes a transition to a re-entrant spin-glass phase around 40 K. The low temperature phase possesses the dynamic characteristics of a re-entrant spin-glass and the ferromagnetic phase has the dynamics of a frustrated ferromagnet, in addition, the system exhibits glassy dielectric properties all of which are illustrated in Fig. 9. The fact that this $La_2NiMnO_6$ sample, a multi-glass, possesses strong magneto-dielectric coupling is noteworthy from fundamental as well as applied points of view.

## 8. Comments
There is an extensive body of literature on the properties of magnetic materials exhibiting competing interaction and disorder. Many of the articles comprising this body concern, however, esoteric realisations of ideal theoretical models. In the quest for new materials with properties that improve on the existing ones or inspire to create new appliances, magnetic materials with competing exchange interactions are prominent candidates. When describing and interpreting empirical findings on such systems, spin-glass and related concepts are frequently used to clarify the nature and structure of complex correlated magnetic states. Certain of these physical concepts are especially appealing, but also deceptive:

**Linear response** – intrinsic dynamics of glassy magnets are observed in regimes of linear response to a field application-removal [23,34]. Experimental studies of glassy magnets thus require that linear response is certified to yield data that allow reliable conclusions and interpretations in terms of spin-glass like features.

**Critical slowing down/critical behaviour** – on approaching a spin-glass transition, critical slowing down occurs and the dynamic exponent is large enough to allow study of the increasing relaxation time on time scales available in standard ac-susceptometers. The analysis requires adequate criteria to determine the freezing temperature and data in a wide enough frequency range to certify that critical slowing down occurs. Examining the literature, few examples of truly reliable analyses are found, however, $zv = 9.5\pm1$ contains conceivable values for 3d Ising spin-glasses and superspin-glasses [5, 25]. Incidentally, too many reported values of critical exponents and parameters related to phase transitions in frustrated magnets are unfortunately ill-derived and given with a precision that far exceeds the experimental accuracy.

**Aging/memory/rejuvenation** – aging, as illustrated in Fig. 4 is a very prominent feature of glassy magnetism – it is, however, difficult to measure in standard magnetometers, where the field application time is often of the order of several seconds or even minutes. An experimentally much simpler protocol to distinguish spin-glass-like glassy dynamics would be ac- [12] or dc- [14] memory experiments.

**AT-lines** – the bifurcation points of ZFC- and FC-magnetisation vs. temperature curves at different applied fields form a constant relaxation time contour when plotted in the H-T plane (representing a time of the order of 10 seconds). This line is not an Almeida-Thouless (AT)-line. A solid determination of an AT-line would require verification of critical slowing down and derivation of the spin-glass temperature $T_g$ as a function of applied field. There is not, as yet, a reliable derivation of an AT-line in any magnetic system. Nevertheless, recent MC simulations [21] indicate that there is no AT-line in 3-dimensional spin glasses and it has been solidly demonstrated experimentally that the spin-glass transition is destroyed in a magnetic field in 3-dimensional Ising spin-glasses [26, 35].

A well considered use of spin-glass concepts to better understand the structure and dynamics of frustrated and inhomogeneous magnetic systems is enlightening [36]. However, when these concepts are used to characterise ill-defined glass transitions or magnetic dynamics measured without certified linear response, they cause confusion instead of clarification of the properties of the investigated material.

Finally, returning to the questions in the title: "Competing interaction in magnets: the root of ordered disorder?" The low temperature states of spin assemblies subject to competing interaction appear magnetically disordered on the length scales of diffraction experiments, but hidden order renders the complex macroscopic physical properties of these systems very different from those of the disordered paramagnetic state at high temperatures; "or only frustration?" Quoting J-P Bouchaud et al. from Phys. Rev. B 65, 024439 (2001) [37] "Although spin-glasses are totally useless pieces of material, they constitute an exceptionally convenient laboratory frame for theoretical and experimental investigations".

ACKNOWLEDGEMENTS
Language corrections from Suzy Harris and artistic scientific contributions from Roland Mathieu are thankfully acknowledged. Financial support from the Swedish Research Council (VR) is acknowledged.